\documentclass[twocolumn,fleqn,natbib]{svjour2}
\bibpunct{[}{]}{;}{n}{}{,} 
\smartqed  
\usepackage{graphicx}

\usepackage{color}

\journalname{Granular Matter}

\begin{document}

\title{On the use of magnetic particles to enhance the flow of vibrated grains through narrow apertures}

\author{C. Manuel Carlevaro \and Marcelo N. Kuperman \and Sebasti\'an Bouzat \and Luis A. Pugnaloni \and Marcos A. Madrid 
} \institute{C. Manuel Carlevaro: Instituto de F\'isica de L\'iquidos y Sistemas Biol\'ogicos, CONICET, 59 789, 1900 La Plata, Argentina
and Departamento de Ingenier\'ia Mec\'anica, Universidad Tecnol\'ogica Nacional, Facultad Regional La Plata, La Plata, 1900, Argentina.
\\
Marcelo Kuperman: Consejo Nacional de Investigaciones Cient\'{\i}ficas y T\'ecnicas, Centro At\'omico Bariloche (CNEA), (8400) Bariloche, R\'{\i}o Negro, Argentina.
\\
Sebasti\'an Bouzat: Consejo Nacional de Investigaciones Cient\'{\i}ficas y T\'ecnicas, Centro At\'omico Bariloche (CNEA), (8400) Bariloche, R\'{\i}o Negro, Argentina.
\\
Luis A. Pugnaloni: Departamento de F\'isica, Facultad de Ciencias Exactas y Naturales, Universidad Nacional de La Pampa, CONICET,
Uruguay 151, 6300 Santa Rosa (La Pampa), Argentina.
\\
Marcos A. Madrid: Instituto de F\'isica de L\'iquidos y Sistemas Biol\'ogicos, CONICET, 59 789, 1900 La Plata, Argentina
and Departamento de Ingenier\'ia Mec\'anica, Universidad Tecnol\'ogica Nacional, Facultad Regional La Plata, La Plata, 1900, Argentina. E-mail: marcosamadrid@gmail.com
}

\authorrunning{C. M. Carlevaro et al.} 

\date{}

%
%
%
%
%
\maketitle

\begin{abstract}
The flow of grains through narrow apertures posses an extraordinary challenge: clogging. Strategies to alleviate the effect of clogging, such as the use of external vibration, are always part of the design of machinery for the handling of bulk materials. It has recently been shown that one way to reduce clogging is to use a small fraction of small particles as an additive. Besides, several works reported that self-repelling magnetic grains can flow through narrow apertures with little clogging, which suggest these are excellent candidates as ``lubricating'' additives for other granular materials. In this work, we study the effect of adding self-repelling magnetic particles to a sample of grains in two-dimensions. We find that, in contrast with intuition, the added magnetic grains not necessarily aid the flow of the original species.
\end{abstract}

\section{Introduction}
The { interest in the dynamics of granular materials flowing out of a container is not new. The pioneering work of Pierre Huber-Burnand \cite{burnand} in the late 19th century already studied the flow of sand grains within vertical cylinders. Since then, these systems have been drawing the attention of a steadily growing group of researchers, giving rise to an extensive literature and a collection of very exciting results. 
However, a complete physical description and understanding of the phenomenology of granular materials is still far from being achieved \cite{campbell06,carpinlioglu16}.
While initial advances in the knowledge of these systems were linked to the need to explain phenomena observed in silos these studies permeated many aspects of industrial activities. Still, the discharge of grains from a silo \cite{beverloo61,mankoc07,duran00} remains one of the most widely studied problems in the area. In particular, the phenomenon responsible for the formation of clogs that interrupt flow is still matter of debate (see \cite{zuriguel14} and references therein). 

Currently, the dynamics of the flow of particles that are discharged from silos through large holes at the bottom are known in detail. The grains flow freely and without interruptions and the flow rate follows the empirical Beverloo equation (or some of its variants) \cite{franklin55,beverloo61,nedderman82,mankoc07}. Recently, a differential equation for the flow rate ---which is consistent with the Beverloo equation--- has been derived from energy balance providing a first principles explanation for the phenomenon \cite{darias2020}. However, when the hole size is small, the discharge rate decreases and its dynamics becomes much more complex \cite{nedderman82,anand08}. Below a certain outlet diameter, the particles form structurally stable arches that block the outlet \cite{to01,zuriguel05} and that can only be destabilized by external forcing. The strategies and mechanisms that ensure the success in the destabilization of these clogging structures has become just as interesting as the flow of the particles whose movement they prevent \cite{takahashi68,suzuki68,lindemann00,chen06,kumar20}.

The vast majority of the solutions to deal with undesirable clogs consist in active mechanical procedures such as vibrating the setup \cite{hunt99,wassgren02,mankoc09,janda09,lozano12,zuriguel17,guerrero18,guerrero19}, making the exit oscillate \cite{to17} or blowing through the opening \cite{zuriguel05}. Passive mechanisms such as placing an obstacle in front of the exit can be used to drastically reduce the clogging probability without applying vibrations \cite{zuriguel11,endo17}. However, once a clog occurs, external perturbations are still required to resume the flow. 

Some experiments with non-circular or non-spherical particles show that analogies can be traced with spherical particles when defining an effective particle radius \cite{borzonyi16,ashour16,szabo18,Goldberg2018}. This is also true for particles that are soft \cite{ashour17,harth20,hong17,wang21}. However, clogging is much less likely for soft particles. Unfortunately, most of the considered soft particles are also frictionless, which makes unclear whether the reduced clogging is due to softness or lack of friction. Other particles which clog very little are repulsive magnetic grains in two-dimensions (2D) \cite{hernandez17,lumay15,Thorens2021}.

The existence of particles that are less prone to clogging suggests that these could be used as additives in a system of particles that clog often to ease the flow. In a recent work, the addition of particles that are smaller than the ones of the granular material of interest has shown to be effective in increasing the flow for vibrated silos \cite{madrid21}. This is mainly due to the fact that arches that contain a small particle are less stable. Also, it has been suggested that the inclusion of a small fraction of self repulsive magnetic particles can also affect the formation of stable blocking structures \cite{nicolas18}. Similar ideas have been explored in \cite{wang21}, where the authors study the effect of doping a  monodisperse hydrogel sphere ensemble with hard frictional particles of the same size and weight. The authors show that the addition of even a small portion of rigid particles to hydrogel sphere ensembles has a remarkable effect on silo discharge behavior. 

While the lubricating effect due to the addition of small grains to a sample has been elucidated \cite{madrid21}, we now aim at pursuing the idea that the presence of a magnetic repulsion between the grains of the added species can enhance the effect. This conjecture is based on the fact that the repelling magnetic field inhibits the contact between magnetic particles and can lead to localized patches of contact-less particles, and so alter the structural stability of the arches responsible for the jamming.

In this work, we study the flow of binary mixtures of grains through a narrow aperture, in which one of the species has a smaller particle size and also a magnetic dipole that leads to a repulsive interaction. The grains are enclosed in a vibrated two-dimensional hopper. We use Discrete Element Method (DEM) simulations for our study. We find that, in contrast with intuition, the presence of the repelling species does not reduce the stability of clogging arches in most cases. On the contrary, they tend to cancel the lubrication effect achieved by the smaller size of the added grains.

\section{Model and simulations}

\begin{figure}
\centering
\includegraphics[width=0.5\columnwidth]{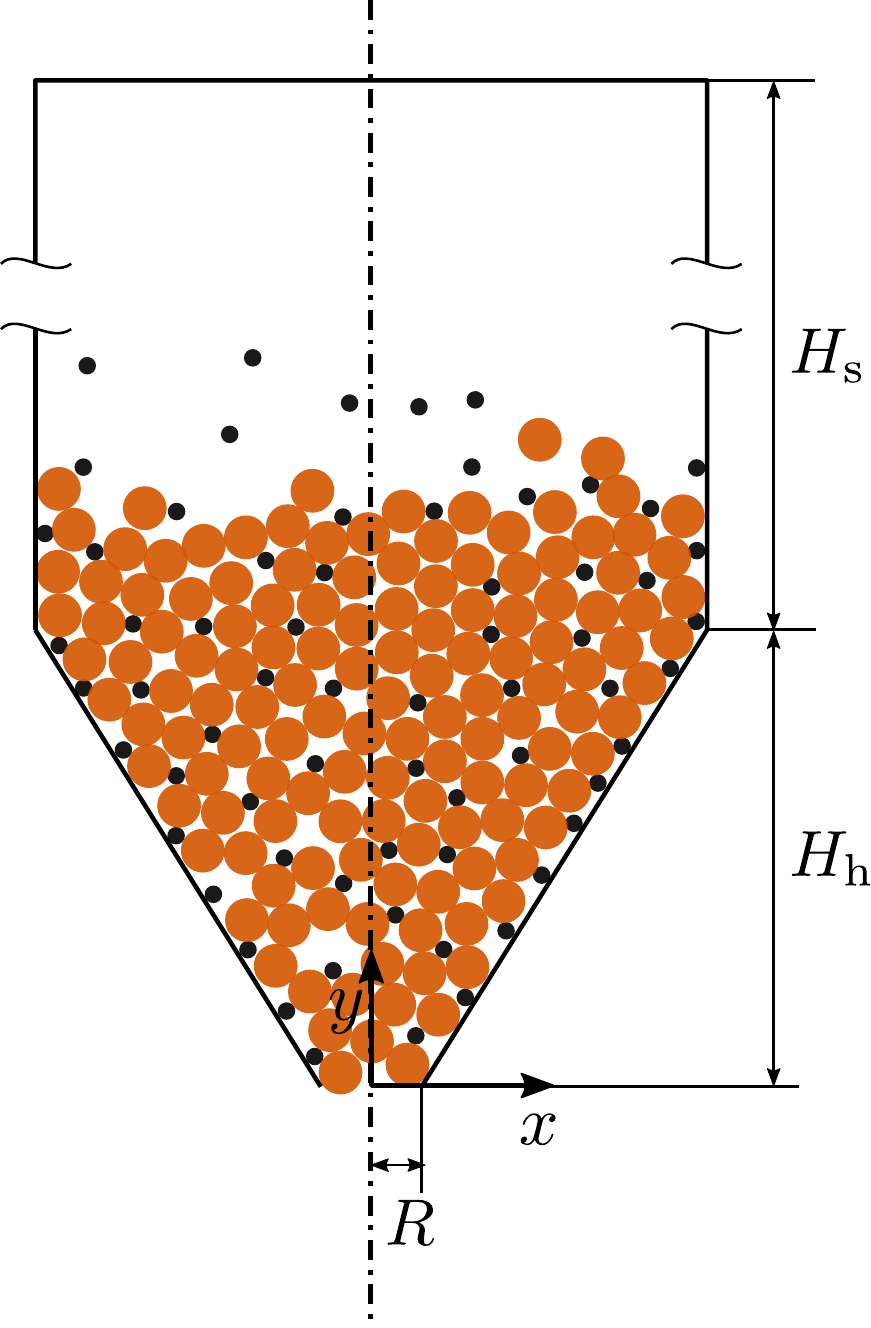}
\caption{\label{scheme} Sketch of the silo and hopper. The red circles correspond to the original species and the blue circles to the added particles.}
\end{figure} 


The numerical simulations use Box2D for the implementation of the DEM \cite{box2d,Catto}. Box2D is a collection of algorithms that has proven to be suited for simulating the dynamics of granular flow of hard grains in 2D under different scenarios  \cite{Goldberg2015,Goldberg2018,Pugnaloni2016,Irastorza2013,Sanchez2014,Carlevaro2020,Pytlos2015}. Before each DEM time step, Box2D utilizes a series of iterations to resolve constraints on overlaps and on static friction between bodies using a Lagrange multiplier scheme~\cite{Pytlos2015, Catto}. This allows to calculate the force at each detected contact considering the Coulomb criterion for the given friction coefficient (static and dynamic friction are set equal) and the restitution coefficient. This simulation scheme is different from traditional event-driven simulations of hard particles in which contacts are only instantaneous and collisions are resolved pair-wise. In Box2D contacts may last many time steps as in soft-particle DEM simulations; however there are not overlaps between object.

Our model considers two types of particles, both of them disks, but with different properties. What we call ``original grains'' are disks of radius $r_o=d/2=6.5$ cm and material density $1.0$ kg/m$^2$. The ``added species''  consists of disks of radius $r_a$, with the same density as the main species and carrying a dipolar magnetic moment $\mu$ perpendicular to the plane on which the 2D particles live. Throughout all the simulations we will set $r_a$ in the range  $r_o/5 < r_a < r_o$.  All the disks will be placed in a 2D vertical silo, subject to vibrations as will be discussed later.
The total number of original and added grains in the silo are referred to as $N_o$ and $N_a$, respectively, and $N_o+N_a=250$ is kept fixed.  The static and dynamic friction coefficients between particles (whatever the species) and between particles and walls are set to $\nu=0.5$. The restitution coefficient is set to $0.1$. The magnetic moment $\mu$ of the added particles leads to a repulsive force only between added grains given by: $ \vec{F} = \frac{3 \mu_0 \, \mu^2}{4 \pi |\vec{r}|^5} \vec{r}$, where $\mu_0 = 4 \pi \times 10^{-7}$ H$\cdot$m$^{-1}$, and $\vec{r}$ is the vector that connects the center of two given magnetic particles.

The vertical silo-and-hopper's total height is $H_s + H_t = 636$ cm $\approx 97.8 r_o$, width of $100$ cm $\simeq 15.3 r_o$ and aperture of radius $R=2.3 r_o$. A scheme of this setup is shown in Fig.~\ref{scheme}. The hopper height $H_t = 136$ cm leads to an angle $(\pi/6)$ between the hopper walls and the vertical direction. The acceleration of gravity is $g=9.8$ m/s$^2$.  

Particles are initially randomly deposited in the 2D silo before opening its aperture at the bottom. In order to maintain the number of particles constant throughout the whole simulation and to avoid spurious effects due to non-stationary states, each particle that crosses the aperture is re-injected above the granular column. To simulate a background vibration that destabilizes the clogs, every $0.1$ s, an individual kick in a random direction (uniformly distributed in [$0,2\pi$]) with an impulse uniformly distributed in the interval [$0,5\times10^{-5}$ Ns] is applied to each particle. At the same time, a global random kick with the same properties as the individual random kicks is applied to all the particles. 

It is important to note that to avoid excessively long clogs, the maximum duration of any clog is limited to $5$ s. If no particle flows through the aperture during $5$ s, the particles that form the clogging arch are identified, removed from the system and re-injected at the top of the granular column. This resumes the flow.

\section{Results}

We consider the effect of two parameters of the problem: the magnetic moment of the dipoles $\mu$ and the size ratio of the added species $r=r_a/r_o$. In all simulations we keep the mixing ratio $\chi=0.4$. For each choice of $\mu$ and $r$ we have carried out $50$ independent realizations of the simulation with different initial random positions of the grains. Each simulation corresponds to $150$ s of discharge.

Introducing particles of small size to a sample of large particles reduces the effective mean particle size. This, in turn, increases the flow rate $Q$ through an orifice of given size and would reduce clogging overall. However, the main point here is to assess the effect on the flow of the original species. Therefore, we focus our attention on the flow rate $\tilde{Q}$ of the original particles only. It is important to bear in mind that the added grains flow along with the original particles and thus take a portion of the time necessary to discharge the mixed system. It is not trivial that the flow rate $\tilde{Q}$ of the original particles be enhanced even if the flow rate of the mixed system is higher than that of the original pure system (which we call $Q_{\rm p} $).

At this point it is worth mentioning that the flux of systems subject to clogging may lead to undefined flow rates if clogs arbitrarily long are likely \cite{zuriguel14-b}. This is of course avoided by using a mechanism to limit the longest clog. Here, the flow rates $Q$ and $\tilde{Q}$ are strongly determined by the choice of waiting time to break long lasting clogs. As mentioned before, we have chosen this cutoff to be $5$ s. Nevertheleyss, if the cutoff is maintained constant, the calculated effective flow rate still serves as a simple parameter to compare the flow between different systems. A measure that characterizes the clogging dynamics and that is not sensitive to this waiting time is the survival function that we present at the end of this section.  

\begin{figure}
	\centering
	\includegraphics[width=0.8\columnwidth]{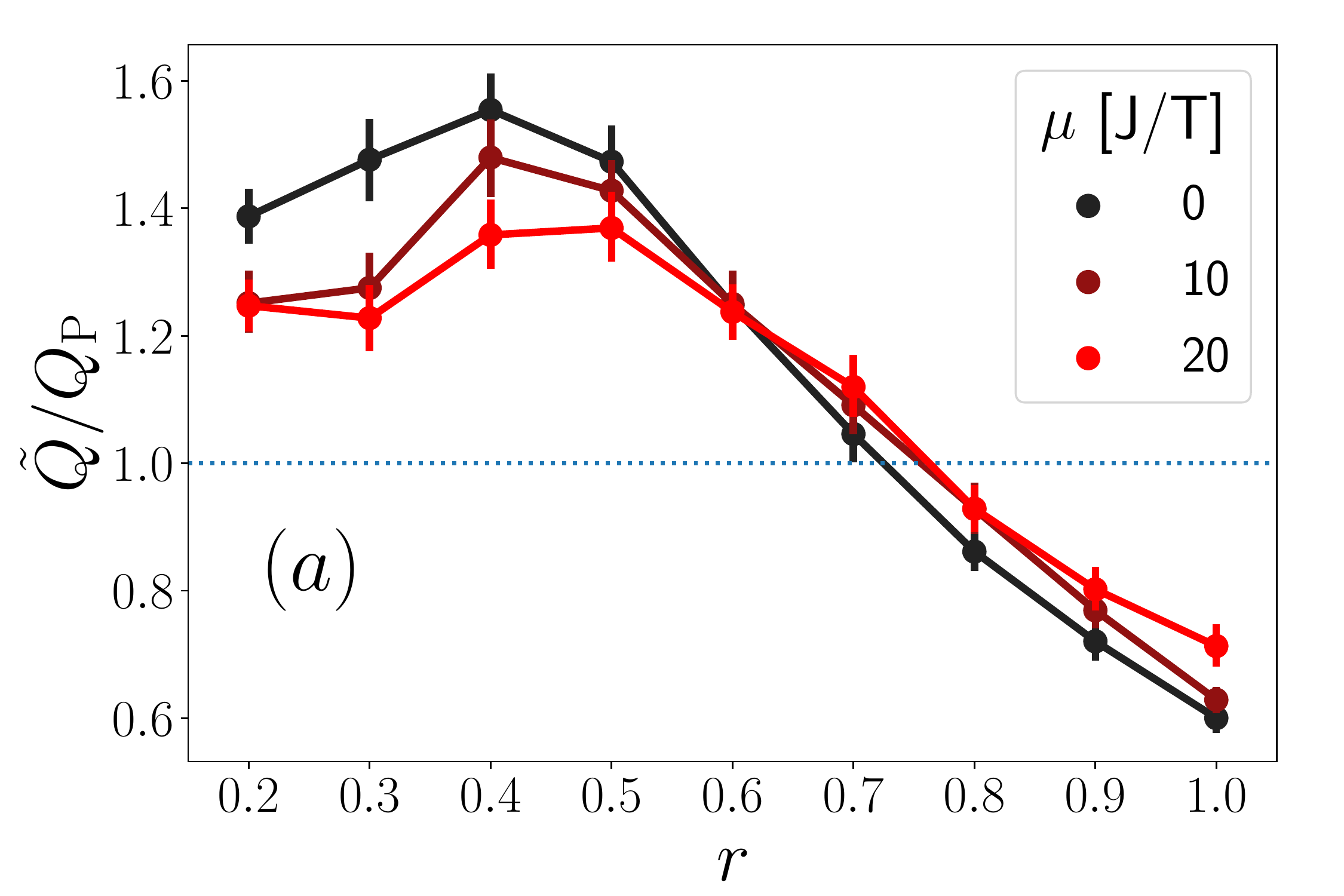}
	\includegraphics[width=0.8\columnwidth]{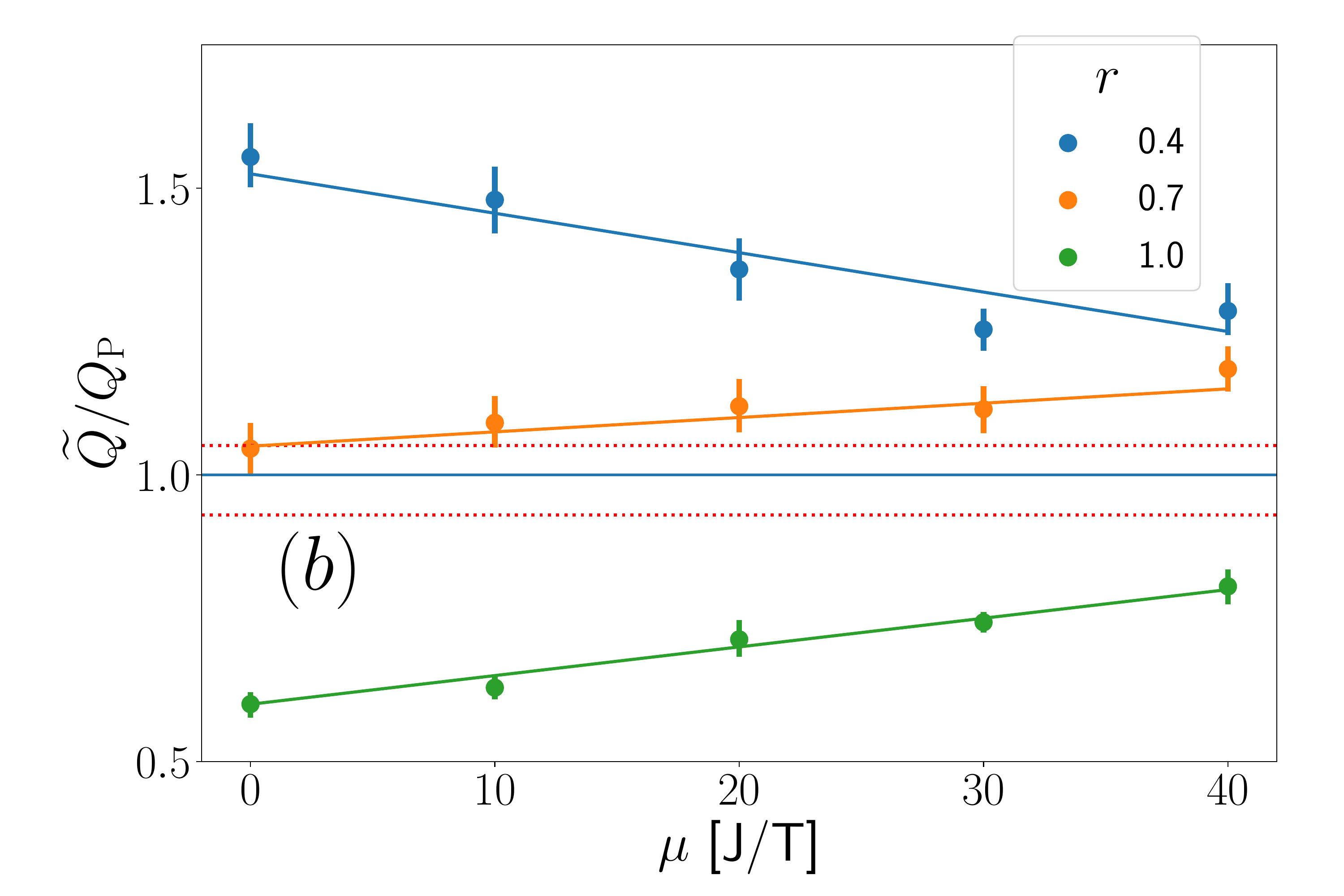}
	\caption{\label{fig3} (a) Flow rate $\tilde{Q}/Q_{\rm p}$ of the original particles as a function $r$ for different values of $\mu$ at $\chi=0.4$. (b) $\tilde{Q}/Q_{\rm p}$ as a function of $\mu$ for $\chi=0.4$ and three different values of $r$. The symbols correspond to the mean values over 50 realizations while bars correspond to the $95\%$ confidence interval.}
	\label{fig2}
\end{figure}

In \cite{madrid21} we have shown that, for nonmagnetic added grains, $\tilde{Q}$ may be higher or lower than $Q_{\rm p}$ depending on the relative size of the particles [see blue data in Fig.~\ref{fig2}(a)]. For $r>0.7$ we find that $\tilde{Q}<Q_{\rm p}$, whereas for $r<0.7$ we find that $\tilde{Q}>Q_{\rm p}$. Moreover, there exists a maximum in $\tilde{Q}$ at $r\approx 0.4$. We have explained these results based on the competition of three effects \cite{madrid21}. Firstly, the added particles reduce the flow rate $\tilde{Q}$ of the originals since some time has to be used to discharge the additional gains (this extra time decreases as $r$ decreases). Secondly, the transient clogging arches that contain a small particle are less stable and break sooner increasing the flow rate $Q$ overall (this is more and more effective as $r$ decreases). Finally, the probability that an added particle becomes part of a clogging arch decreases with its size. As a result, adding particles with $r>0.7$ reduces $\tilde{Q}$ (i.e., $\tilde{Q}<Q_{\rm p}$) since the extra time required to evacuate the added particles dominate. For $r<0.7$ the more unstable arches and the lower time required to evacuate smaller grains lead to $\tilde{Q}>Q_{\rm p}$. However, for $r<0.4$ the low probability that a small particle be involved in an arch makes $\tilde{Q}$ to decline and tend to $Q_{\rm p}$.   

Figure~\ref{fig2}(a) includes results for repulsive added particles. As we can see, the inclusion of a magnetic dipole leads to a second order effect on top of the general trends dominated by the relative particle size. For $r> 0.6$, $\tilde{Q}$ grows with the magnetic repulsion. In contrast, for $r<0.6$, $\tilde{Q}$ decreases with $\mu$. To highlight these contrasting behaviors we plot in Fig. \ref{fig2}(b) $\tilde{Q}$ as a function of $\mu$ for some selected size ratios. For added particles of the same size as the original particles ($r=1.0$) the flow rate of the originals is never greater than $Q_{\rm p}$. The magnetic repulsion does help in improving $\tilde{Q}$ with respect to nonmagnetic added grains but this is not able to overcome the reduction in the flow of the original species caused by the fact that the added particles take a significant portion of time to be evacuated. When we focus on small added particles ($r=0.4$), the gain in $\tilde{Q}$ is reduced by the presence of the magnetic repulsion, which sound, in principle,  counterintuitive. Finally, we observe a narrow range of particle sizes ($0.6<r<0.7$) in which the flow rate of the originals is improved by the nonmagnetic added grains and the addition of magnetic repulsion provides an additional subtle increase in $\tilde{Q}$.

The surprising behavior described above can be understood in terms of an interplay between local concentration (at the outlet) of added particles and arch stability. As we mentioned above, small added particles tend to reduce the stability of the arches in which they intervene. Magnetic repulsion between the added grains keeps them apart (see movie in Supplementary material \cite{supplementary1}) reducing the frequency at which small particles are involved in arches to ease the flow. This is why for relative small added grains ($r<0.6$) we observe a decrease of $\tilde{Q}$ with $\mu$. For $r>0.6$, however, the reduction in arch stability is less marked and the time used to evacuate added grains plays against the flow rate of the original particles. The inclusion of a magnetic repulsion between the added particles in this case serves to reduce the local density of these and create a lower effective $\chi$, which improves the flow rate $\tilde{Q}$. We note that at the crossover ($r\approx 0.6$) the effect of $\mu$ is negligible. However, the precise location of this crossover may depend on $\chi$ and on other parameters including the background noise.

\begin{figure}
\centering
 \includegraphics[width=0.8\columnwidth]{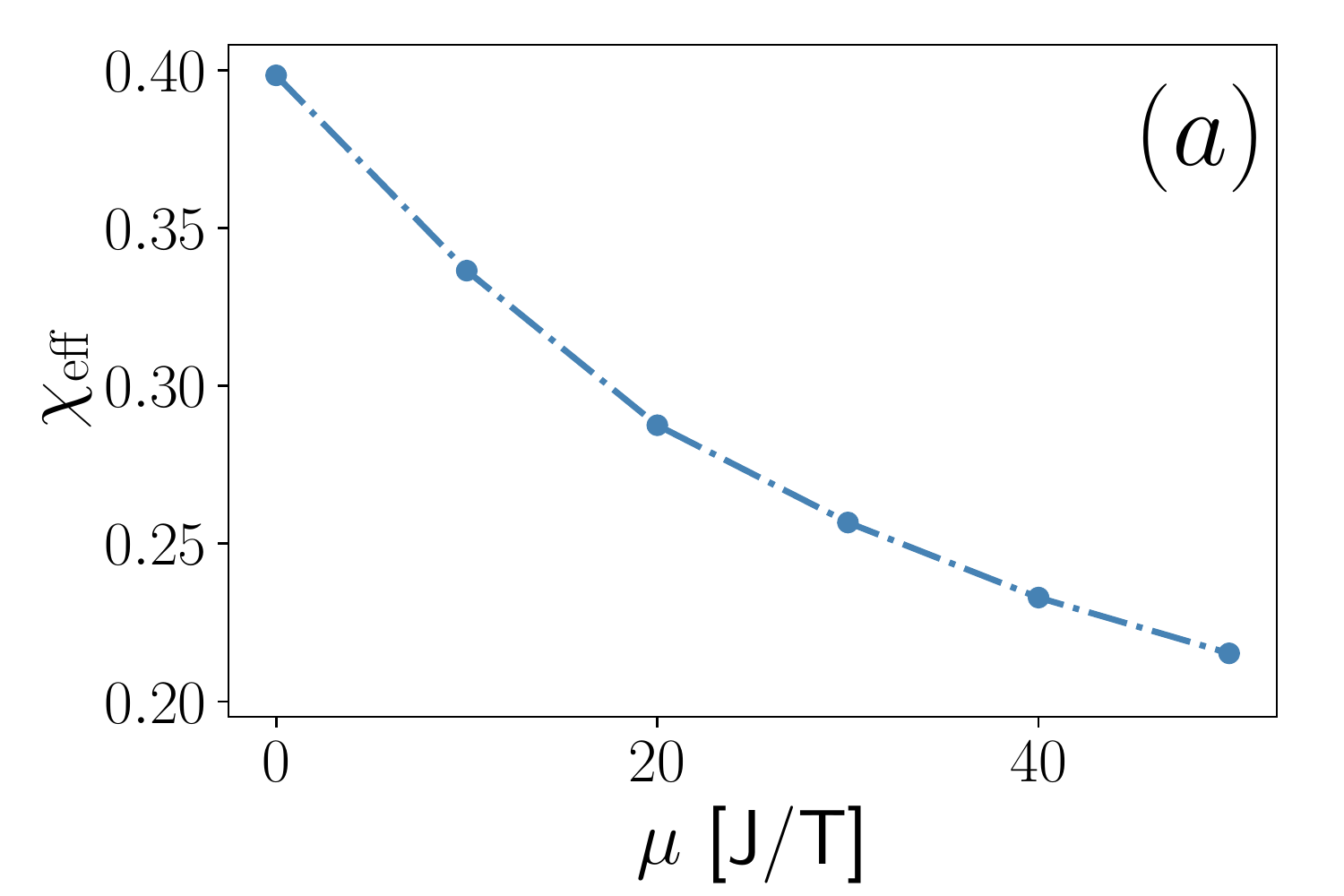}
 \includegraphics[width=0.8\columnwidth]{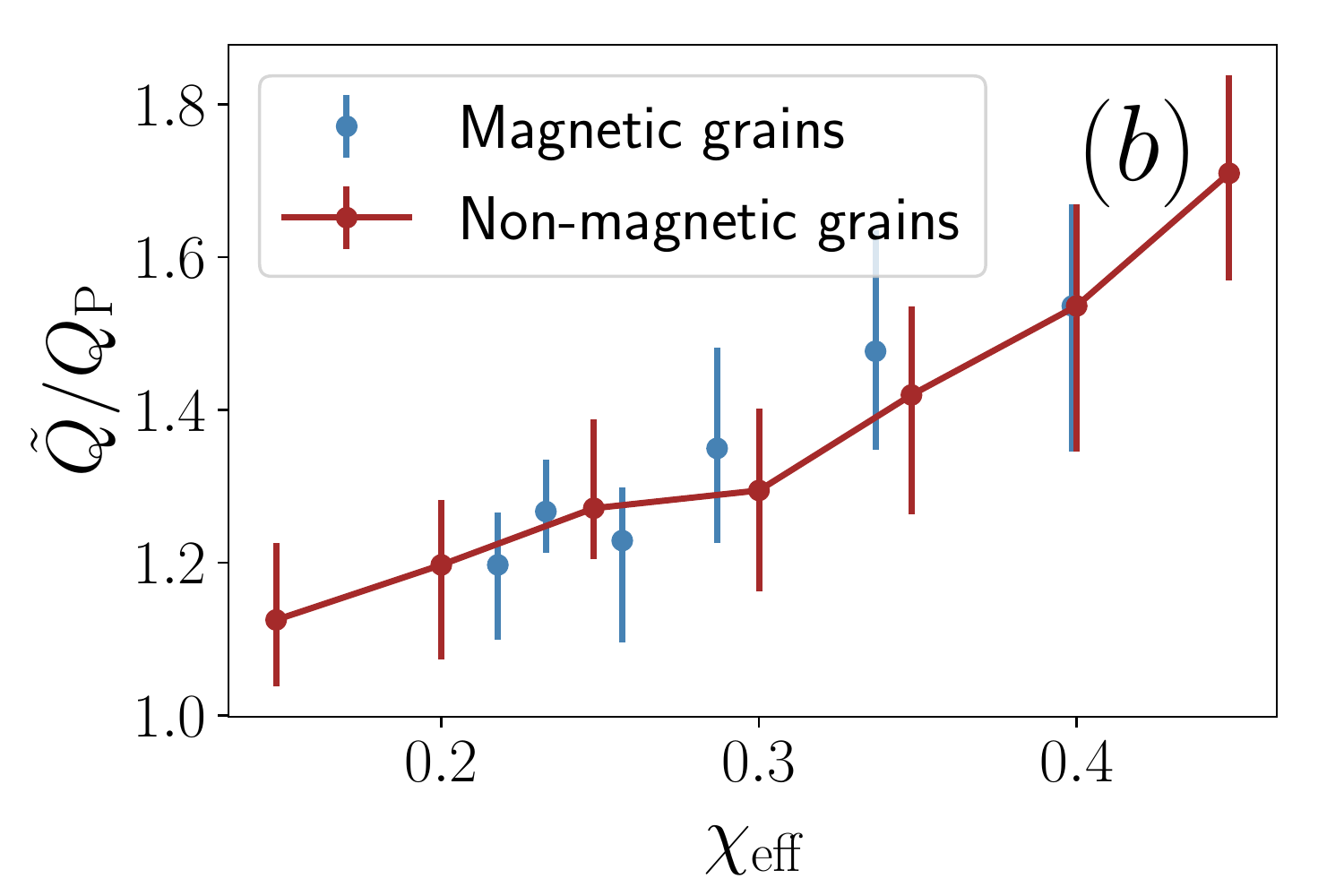}
 \caption{\label{fig:chi}
(a) Effective mix ratio $\chi_{\rm eff}$ at the outlet as a function of $\mu$. (b) $\tilde{Q}$ as a function of $\chi_{\rm eff}$ for magnetic added particles, compared with $\tilde{Q}$ for nonmagnetic added particle as a function of $\chi$.}
\end{figure}

We have confirmed the reduction of the effective mix ratio with increasing $\mu$ by measuring the mix ratio $\chi_{\rm eff}$ of the outflowing grains. In Fig.~\ref{fig:chi}(a) we plot $\chi_{\rm eff}$ as a function of $\mu$. As we can see, $\chi_{\rm eff}(\mu)$ decreases with $\mu$ indicating that fewer magnetic added particles are present at the orifice as $\mu$ grows. In fact, many of the magnetic grains can be observed floating on top of the packed column of grains (see Supplemental material \cite{supplementary1}). In Fig.~\ref{fig:chi}(b) we plot $\tilde{Q}$ as a function of $\chi_{\rm eff}$ for magnetic added particles. For comparison, we also plot $\tilde{Q}$ for nonmagnetic added particles where we have changed the actual value of $\chi$. The two curves agree rather well. This demonstrates that, to leading order, the magnetic repulsion has an effect equivalent to reducing the local concentration of added particles. This effective reduction of $\chi$ diminishes the effect of the corresponding nonmagnetic case, whether it was an enhancement or a deterioration of the effective flow rate with respect to the pure system.

We obtain additional information about the clogging dynamics by  exploring the survival function $\tilde{P}(\Delta t\geq\tau)$, which gives the probability that we find a time gap $\Delta t > \tau$ between the egresses of two consecutive original particles \cite{nicolas16}. It is important to note that a number of added grains may exit the container between the passage of two original particles. In Fig. \ref{figsur} we compare the survival function for the pure system with those for mixed 
systems considering different $\mu$ for the magnetic added particles (we fixed $r=0.4$ and $\chi=0.4$). We note that $\tilde{P}(\Delta t\geq\tau)$ vanishes for $\Delta t>5$ s due to the mechanism we introduced for breaking long lasting clogs. As we have described in \cite{madrid21}, introducing small nonmagnetic particles reduces significantly the occurrence of long lasting clogs. From there, as $\mu$ is increased, the probability of long lasting clogs increases in agreement with the reduction of the flow rate (see Fig.~\ref{fig2}). However, for the values of $\mu$ explored, the pure system always presents more long lasting clogs.

\begin{figure}
\centering
	\includegraphics[width=\linewidth]{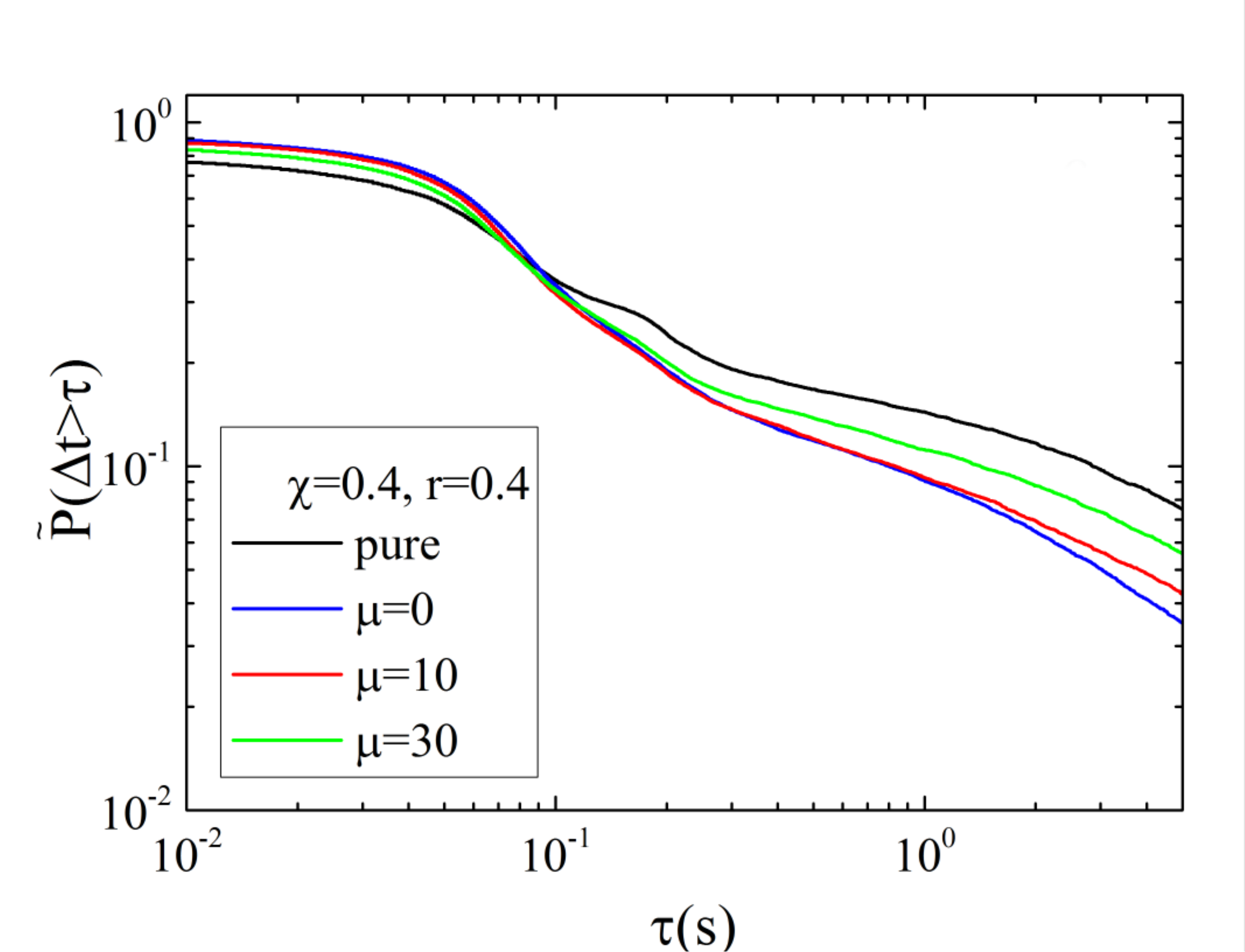}
	\caption{\label{figsur} Survival function for different values of $\chi$ (a),  $r$ and $\mu$ (c). The black solid line corresponds to the pure system with no added particles.}
\end{figure}

As we mentioned, whenever a blocking arch lasts more than $5$ s we remove all the particles that form that arch to resume the flow. We have analyzed the number and composition of these long lasting arches. Figure \ref{fig5}(a) shows the total number of long lasting arches ($\tau > 5$ s) that needed to be removed per minute during any given simulation as a function of $\mu$. We can observe that the smaller the added particles the lower the occurrence of these long lasting arches (in agreement with \cite{madrid21}). However, an increase in $\mu$ leads to a larger number of long lasting arches.

As discussed above, repulsion between the added particles leads to a smaller participation of these added particles in the arches. Therefore, part of their contribution to make arches less stables is lost. To validate this hypothesis we plot in Fig. \ref{fig5}(b) the number of small particles involved in long lasting arches. As expected, the addition of a magnetic repulsion leads to a decrease in the number of small particles in the arches. Naturally, this effect is expected to occur for all arches and not only for long lasting ones. Here, it is important to recall that more long lasting arches (and fewer added particles in these arches) does not always lead to a reduction of $\tilde{Q}$. The portion of time devoted to evacuate the added particles is also an important factor and this time decreases if fewer added particles reach the outlet due to the magnetic repulsion. Fro this reason, if $r>0.6$ the flow rate of the original species is enhanced by the repulsive interaction of the added grains. However, only for $0.6<r<0.7$ this small improvement in $\tilde{Q}$ is in fact above $Q_{\rm p}$.

\begin{figure}
\centering
\includegraphics[width=0.49\columnwidth]{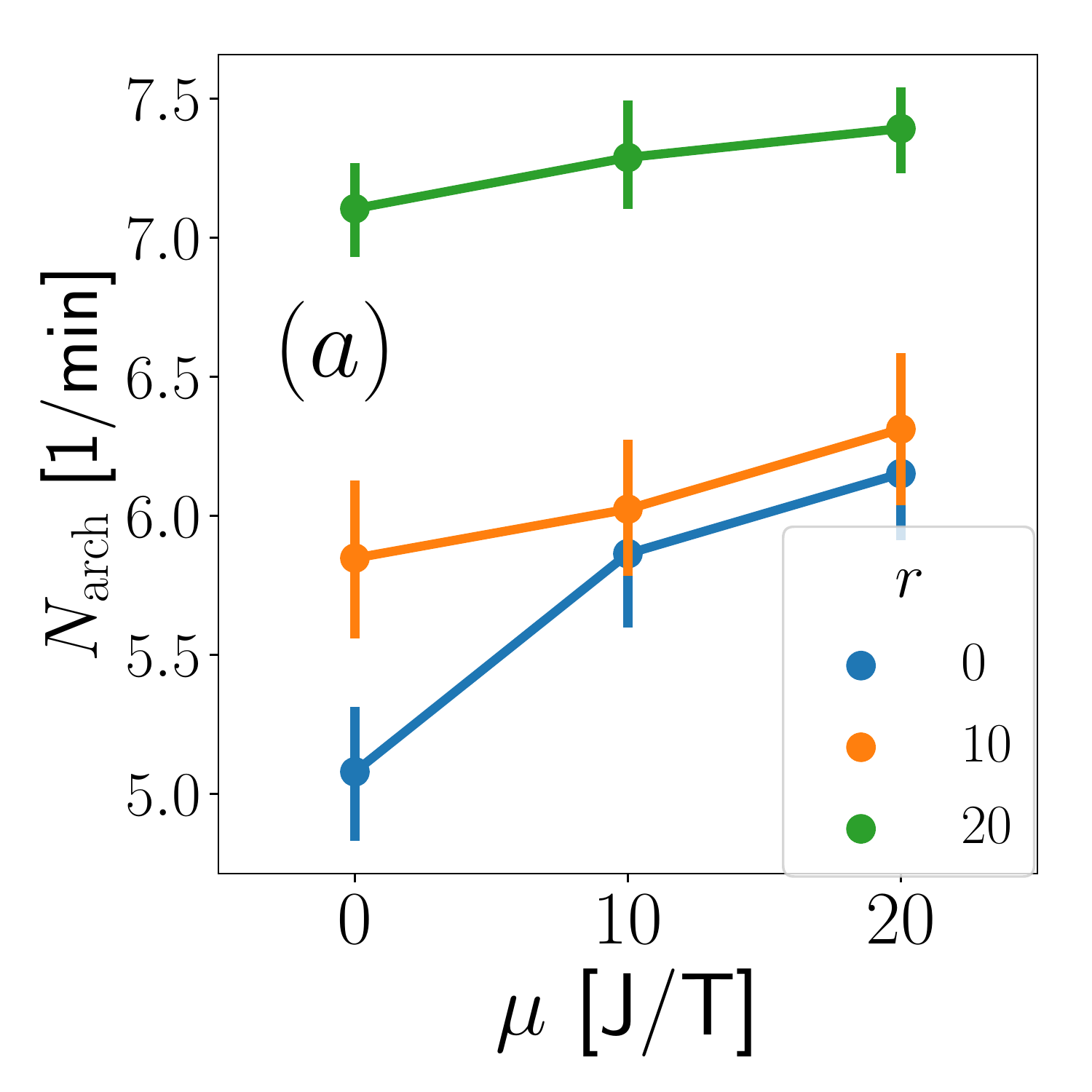}
\includegraphics[width=0.49\columnwidth]{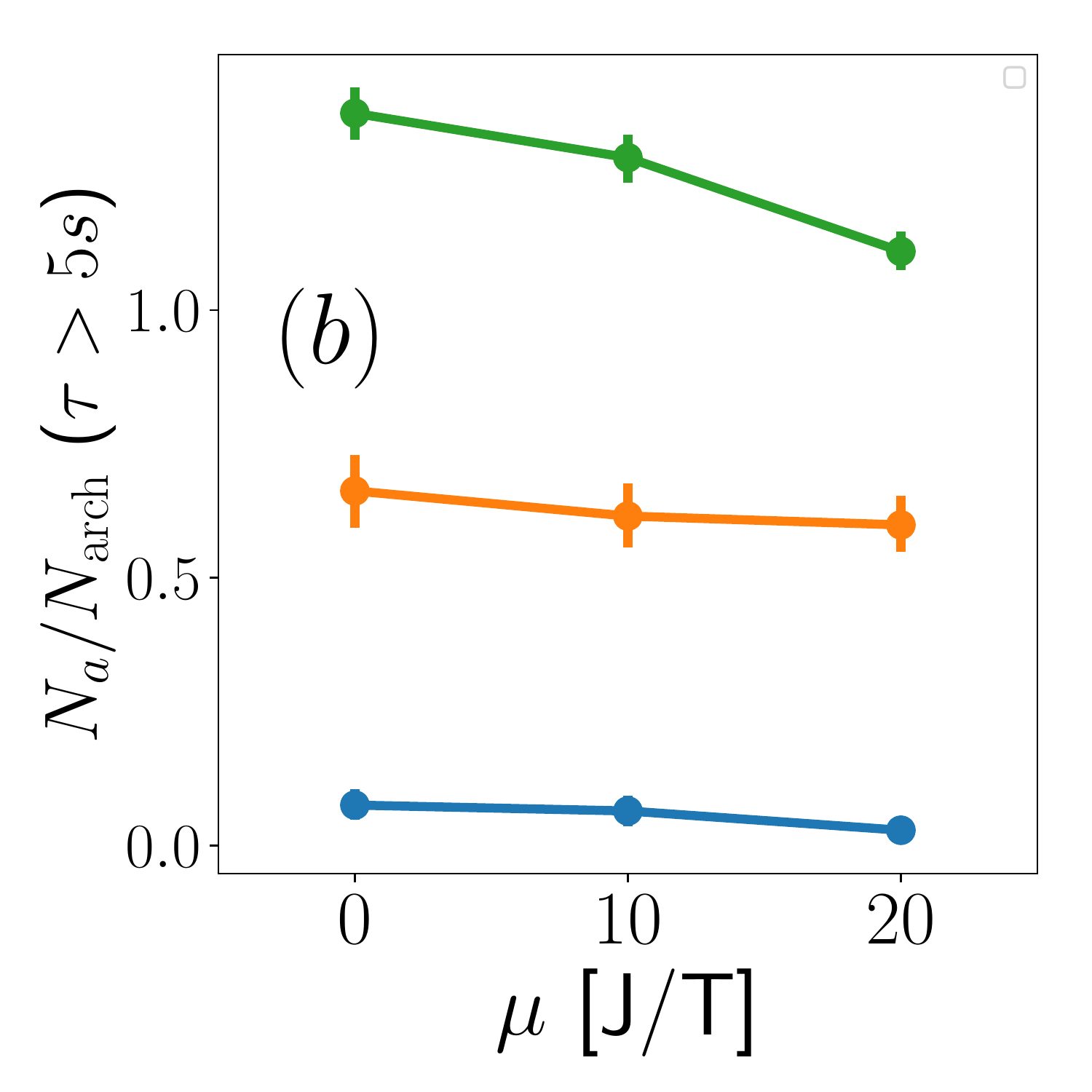}
	\caption{\label{fig5} (a) Number of long lasting arches ($\tau >5$ s) that block the aperture per minute as a function of $\mu$ for different values of $r$ at $\chi=0.4$. (b) Number $N_a$ of small added particles per arch in long lasting arches ($\tau >5$ s) as a function of $\mu$ for different values of $r$ at $\chi=0.4$. The error bars correspond to the standard deviation over $50$ realizations.}
\end{figure}

\section{Conclusions}

In the search for strategies to improve the granular flow when emptying a silo though a small aperture, some ideas appear very intuitive based on the knowledge we have about the behavior of different simple (one single species) granular system. In particular, repulsive magnetic particles have shown a very low clogging probability. Hence, it seems natural to expect that adding some magnetic particles to a nonmagnetic granular system will reduce clogging and improve the effective flow of the original particles. 
In the present work, we studied the effect of a repulsive interaction between the particles of an added species on the flow of the original species. Since we have previously shown that the size of the added particles plays an important role \cite{madrid21}, we have tested different particle sizes. 

We have found that the addition of the repulsive interaction leads to a lower number of added grains in the region of the orifice. As a result, in general, any effect caused by the size of the added particles (either improving or reducing the effective flow of the original grains) is partially suppressed by the repulsive interaction. For the system studied, only in a narrow region of added particle sizes ($0.6<r<0.7$) we observe that the small enhancement of the effective flow is further improved by the magnetic repulsion.

\section*{Acknowledgements} 
The authors would like to thank the financial support from CNEA,
CONICET (through Grants No. PIP 112 2017 0100008 CO and No. PUE 2018 229 20180100010 CO) and
UTN (under Grants No. MAUTNLP0004415 and No. MAUTNLP0006542), all Argentinean public institutions.
 
\section*{Compliance with ethical standards}
Conflict of Interest: The authors declare that they have no conflict of interes.


\begin{thebibliography}{99}

\bibitem{burnand} Huber-Burnand, P.:  Ueber den Ausfluß und den Druck des Sandes, Kunst- und Gewerbeblatt des Polytechnischen Vereins für das Königreich Bayern \textbf{15}, 730–740 (1829).


\bibitem{campbell06} Campbell, C. S.: Granular material flows - An overview, J. Powder Technol. \textbf{162}, 208 (2006).

\bibitem{carpinlioglu16}  Carpinlioglu, M.O.: A critical review on modeling and analysis of granular matter flows, Int. J. Chem.
Chem. Eng. Syst. \textbf{1}, 21 (2016).


\bibitem{duran00} Duran, J., Sands: Powders, and Grains: An Introduction to the Physics of Granular Materials, (Springer-Verlag, New York, 2000).

\bibitem{beverloo61} Beverloo, W.A., Leniger, H. A. and Van de Velde, J. J.: The flow of granular solids through orifices, Chem.Eng. Sci. \textbf{15}, 260-269 (1961).

\bibitem{mankoc07} Mankoc, C., Janda, A.,  Arévalo, R., Pastor, J. M., Zuriguel, I.,
Garcimart{\'\i}n, A., and  Maza, D.: The flow rate of granular materials through an orifice, Granular Matter \textbf{9},
407–414 (2007).

\bibitem{zuriguel14}  Zuriguel, I. : Clogging of granular materials in bottlenecks, Papers in Physics \textbf{6}, 060014 (2014)

\bibitem{franklin55} Franklin, F. C. and  Johanson, L. N: Flow of granular material through a circular orifice, Chem. Eng. Sci. \textbf{4},
119–129 (1955).

\bibitem{nedderman82} Neddermann, R. M., Tuzun, U., Savage, S. B. and Houlsby, G. T.:The flow of granular materials—I: Discharge rates from hoppers, Chem. Eng. Sci.\textbf{37}, 1597–1609 (1982).

\bibitem{darias2020} Darias, J.R., Madrid, M. A. and Pugnaloni, L. A. : Differential equation for the flow rate of discharging silos based on energy balance, Phys. Rev. E \textbf{101}, 052905 (2020).


\bibitem{anand08} Anand, A., Curtis, J. S., Wassgren,  C. R., Hancock,  B. C.  and
Ketterhagen, W. R. : Predicting Discharge Dynamics From A Rectangular Hopper Using The Discrete Element Method (Dem), Chem. Eng. Sci. \textbf{63}, 5821 (2008).

\bibitem{to01} To, K.,  Lai, P. and  Pak, H. K. : Jamming of Granular Flow in a Two-Dimensional Hopper, Phys. Rev. Lett. \textbf{86}, 71 (2001).

\bibitem{zuriguel05} Zuriguel, I., Garcimart\'in, A., Maza, D., Pugnaloni, L. A. and Pastor,  J. M.: Jamming during the discharge of granular matter from a silo, Phys. Rev. E \textbf{71}, 051303 (2005).

\bibitem{takahashi68} Takahashi, H., Suzuki, A. and Tanaka, T.: Behaviour of a particle bed in the field of vibration I. Analysis of particle motion in a vibrating vessel, Powder Technol. \textbf{2}, 65-71 (1968).

\bibitem{suzuki68} Suzuki, A., Takahashi, H. and Tanaka, T.: Behaviour of a particle bed in the field of vibration II. Flow of particles through slits in the bottom of a vibrating vessel, Powder Technol.\textbf{2}, 72-77 (1968).

\bibitem{lindemann00} Lindemann, K. and Dimon, P.: Two-dimensional granular flow in a vibrated small-angle funnel, Phys. Rev. E \textbf{62}, 5420 (2000).

\bibitem{chen06}  Chen, K., Stone, M. B., Barry, R., Lohr, M., McConville, W., Klein, K., Sheu, B. L., Morss, A. J., Scheidemantel, T. and Schiffer, P.: Flux through a hole from a shaken granular medium , Phys. Rev. E \textbf{74}, 011306 (2006).


\bibitem{kumar20} Kumar, R. , Jana, A. K., Gopireddy, S. R. and  Patel, C. M.: Effect of horizontal vibrations on mass flow rate and segregation during hopper discharge: discrete element method approach, S\^adhan\^a \textbf{45}, 67 (2020).


\bibitem{hunt99} Hunt, M. L. and Weathers, R. C. and Lee, A. T. and Brennen, C. E. and Wassgren, C. R:  Effects of horizontal vibration on hopper flows of granular materials , Phys. Fluids \textbf{11}, 68 (1999).


\bibitem{wassgren02} Wassgren, C. R., Hunt, M. L., Freese, P. J., Palamara, J. and Brennen, C. E.: Effects of vertical vibration on hopper flows of granular material, Phys. Fluids \textbf{14}, 3439 (2002).

\bibitem{mankoc09} Mankoc, C., Garcimart{\'\i}n, A., Zuriguel, I., Maza, D. and Pugnaloni, L. A.: Role of vibrations in the jamming and unjamming of grains discharging from  a silo, Phys. Rev. E \textbf{80}(1), 011309 (2009).

\bibitem{janda09} Janda, A., Harich, R., Zuriguel, I., Maza, D., Cixous, P. and  Garcimartin, A.: Flow-rate fluctuations in the outpouring of grains from a two-dimensional silo, Phys. Rev. E \textbf{79}, 031302 (2009).


\bibitem{lozano12} Lozano, C., Lumay, G., Zuriguel, I., Hidalgo, R. and Garcimart{\'\i}n, A.: Breaking arches with vibrations: the role of defects, Phys. Rev. Lett. \textbf{109}(6), 068001 (2012).


\bibitem{zuriguel17} Zuriguel, I., Janda, A., Ar\'evalo, R., Maza, D. and Garcimart{\'\i}n, A.: Clogging and unclogging of many-particle systems passing through a bottleneck, EPJ Web Conf. \textbf{140}, 01002 (2017).


\bibitem{guerrero18} Guerrero, B. V., Pugnaloni, L. A., Lozano, C., Zuriguel, I. and  Garcimart{\'\i}n, A.: Slow relaxation dynamics of clogs in a vibrated granular silo, Phys. Rev. E \textbf{97}, 042904 (2018).

\bibitem{guerrero19} Guerrero, B. V., Chakraborty, B., Zuriguel, I. and , Garcimart{\'\i}n, A.: Nonergodicity in the silo unclogging: Broken and unbroken arches, Phys. Rev. E \textbf{100}, 032901 (2019).


\bibitem{to17} To, K. and  Tai, H.  T.: Flow and clog in a silo with oscillating exit, 	Phys. Rev. \textbf{96}, 032906 (2017).

\bibitem{zuriguel11} Zuriguel, I., Janda, A., Garcimart{\'\i}n, A., Lozano, C., Ar{\'e}valo, R. and
Maza, D.: Silo clogging reduction by the presence of an obstacle, Phys. Rev. Lett. \textbf{107},278001 (2011).

\bibitem{endo17} Endo, K, Reddy, K. A. and Katsuragi, H.: Obstacle-shape effect in a two-dimensional granular silo flow field, Phys. Rev. Fluids \textbf{2},094302, (2017).  

\bibitem{borzonyi16}  B\"orzs\"onyi, T., Somfai, E., Szab\'o, B., Wegner, S., Mier, P., Rose, G. and  Stannarius, R.: Packing, alignment and flow of shape-anisotropic grains in a 3D silo experiment, New J. Phys. \textbf{18}, 093017 (2016).


\bibitem{ashour16} Ashour, A., Wegner, S., Trittel, T., B\"orzs\"onyi, T. and Stannarius, R.: Outflow and clogging of shape-anisotropic grains in hoppers with small apertures, Soft Matter \textbf{13}, 402-414 (2017).

\bibitem{szabo18} Szab\'o, B., Kov\'acs, Z., Wegner, S., Ashour, A., Fischer, D., Stannarius, R., and B\"orzs\"onyi, T.: Flow of anisometric particles in a quasi-2D hopper, Phys. Rev. E \textbf{97}, 062904(2018).


\bibitem{Goldberg2018} Goldberg, E., Carlevaro, C, M. and  Pugnaloni, L. A.:Clogging in two-dimensions: effect of particle shape, J. Stat. Mech. {\bf 2018}, 113201 (2018)


\bibitem{ashour17} Ashour, A., Trittel, T., Börzsönyi, T. and Stannarius, R. :Silo outflow of soft frictionless spheres, Phys. Rev. Fl. \textbf{2}, 123302 (2017)

\bibitem{harth20} Harth, K.,  Wang, J., Börzsönyi, T. and Stannarius, R. :Intermittent flow and transient congestions of soft spheres passing narrow orifices, Soft Matter \textbf{16},8013-8023 (2020). 


\bibitem{hong17} Hong, X., Kohne, M., Morrell, M., Wang,  H. and  Weeks, E. R.: Clogging of soft particles in two-dimensional hoppers, Phys Rev.  E \textbf{96}, 062605 (2017)


\bibitem{wang21} Wang, J., Fan, B., Pong\'o, T., Harth, K., Trittel, T., Stannarius, R., Illig, M., B\"orzs\"onyi  T. and
Cruz Hidalgo, R.: Silo discharge of mixtures of soft and rigid grains, Soft Matter \textbf{17}, 4282-4295 (2021). 


\bibitem{lumay15} Lumay, G., Schockmel, J., Hen{\'a}ndez-Enr{\'\i}quez, D., Dorbolo, S., Vandewalle, N. and F.~Pacheco-Vazquez, F. : Flow of magnetic repelling grains in a two-dimensional silo, Papers in  Physics \textbf{7}, 070013 (2015).


\bibitem{hernandez17} Hernández-Enríquez, D.,  Lumay, G., Pacheco-Vázquez, F.: Discharge of repulsive grains from a silo: experiments and simulations, EPJ Web of Conferences \textbf{140}, 03089 (2017)

\bibitem{Thorens2021} Thorens, L., Viallet, M., Måløy, K. J., Bourgoin, M. and  Santucci, S.: Discharge of a 2D magnetic silo, EPJ WoC {\bf 249}, 03017 (2021).

\bibitem{madrid21} Madrid, M. A., Carlevaro, C. M., Pugnaloni, L. A., Kuperman, M. and  Bouzat, S.: Enhancement of the flow of vibrated grains through narrow apertures by addition of small particles, Phys. Rev. E \textbf{103}, L030901 (2021).


\bibitem{nicolas18} Nicolas, A., Ib\'a\~nez, S., Kuperman, M. N.  and Bouzat, S.: A counterintuitive way to speed up pedestrian and granular bottleneck flows prone to clogging: can 'more' escape faster?, J. Stat. Mech. \textbf{8}, 083403, (2018). 

\bibitem{box2d} Box2d physics engine, {https://www.box2d.org}

\bibitem{Catto} Catto, E.: Iterative dynamics with temporal coherence, 
{https://box2d.org/publications/} (2005).


\bibitem{Goldberg2015} Goldberg, E., Carlevaro,  C. M.  and Pugnaloni, L. A.: Flow rate of polygonal grains through a bottleneck: Interplay between shape and size, Papers in Physics {\bf 7}, 070016 (2015).


\bibitem{Pugnaloni2016} Pugnaloni, L. A., Carlevaro,  C. M., Kram\'ar, M., Mischaikow, K. and Kondic, L.: Structure of force networks in tapped particulate systems of disks and pentagons. I. Clusters and loops, Phys. Rev. E {\bf 93}, 062902 (2016).

\bibitem{Irastorza2013} Irastorza, R. M., Carlevaro, C. M. and Pugnaloni, L. A. : Exact predictions from the Edwards ensemble versus realistic simulations of tapped narrow two-dimensional granular columns, J. Stat. Mech. {\bf 2013}, P12012 (2013).

\bibitem{Sanchez2014} S\'anchez, M., Carlevaro, C. M. and Pugnaloni, L. A. : Effect of particle shape and fragmentation on the response of particle dampers, J. Vibr. Contr. {\bf 20}, 1846 (2014).

\bibitem{Carlevaro2020} Carlevaro, C. M. , Kozlowski, R., Pugnaloni, L. A., Zheng, H., Socolar,  J. E. S.  and Kondic, L.: Intruder in a two-dimensional granular system: Effects of dynamic and static basal friction on stick-slip and clogging dynamics,  Phys. Rev. E {\bf 101}, 012909 (2020).

\bibitem{Pytlos2015} Pytlos, M., Gilbert, M. and Smith, C. C.: Modelling granular soil behaviour using a physics engine, Geotech. Lett. {\bf 5}, 243 (2015).

\bibitem{nicolas16} Nicolas, A., Bouzat, S. and Kuperman, M. N.: Statistical fluctuations in pedestrian evacuation times and the effect of social contagion, Phys. Rev. E {\bf 94}, 022313 (2016)

\bibitem{zuriguel14-b} Zuriguel, I., Parisi, D. R., Hidalgo, R. C., Lozano, C., Janda, A., Gago, P. A., Peralta, J. P., Ferrer, L. M., Pugnaloni, L. A., Clément, E., Clogging transition of many-particle systems owing through bottlenecks, Sci.Rep. \textbf{4}, 7324 (2014).

\bibitem{supplementary1} A video of the discharge for magnetic and nonmagnetic added particles [\textbf{insert link}].


\end{thebibliography}
\end{document}